\documentclass[aps,prl,twocolumn,super,comma]{revtex4-2}

\usepackage{blindtext,amssymb,gensymb,siunitx,textgreek,textcomp,mathtools,bm,soul}
\usepackage{wasysym} 
\usepackage[version=3]{mhchem} 

\usepackage[utf8]{inputenc}
\usepackage[T1]{fontenc}
\usepackage [english]{babel}
\usepackage [autostyle, english = american]{csquotes}
\MakeOuterQuote{"}
\usepackage{subfiles}
\usepackage{fancyvrb}
\usepackage{lipsum}

\usepackage[margin=0.625in]{geometry}

\usepackage{caption,graphicx}
\usepackage{tablefootnote}

\usepackage{booktabs}
\usepackage[table,xcdraw]{xcolor}
\usepackage[hidelinks]{hyperref}
\usepackage{cleveref,xr}
\setcitestyle{super}

\begin{document}

\title{A method to computationally screen for tunable properties of crystalline alloys}
\author{Rachel Woods-Robinson*\textsuperscript{†}\textsuperscript{1,2}}
\author{Matthew K. Horton*\textsuperscript{2,3}}
\author{Kristin A. Persson\textsuperscript{3,4}
}

\affiliation{\textsuperscript{1}Applied Science and Technology Graduate Group, University of California at Berkeley, Berkeley, CA, 94720 USA, \textsuperscript{2}Materials Sciences Division, Lawrence Berkeley National Laboratory, Berkeley, CA, 94720 USA,
\textsuperscript{3}Department of Materials Science and Engineering, University of California at Berkeley, Berkeley, CA, 94720 USA,
\textsuperscript{4}Molecular Foundry Division, Lawrence Berkeley National Laboratory, Berkeley, CA, 94720 USA, \
*These authors contributed equally to this research. \textsuperscript{†}Lead Contact: rwoodsrobinson@berkeley.edu}

\date{\today}

\begin{abstract}


Conventionally, high-throughput computational materials searches start from an input set of bulk compounds extracted from material databases, but in contrast many real functional materials are heavily-engineered mixtures of compounds rather than single bulk compounds. We present a framework and open-source code to automatically construct and analyze possible alloys and solid solutions from a set of existing, experimental or calculated ordered compounds, without requiring additional metadata except crystal structure. As a demonstration, we apply this framework to all compounds in the Materials Project to create a new, publicly available database of $>$600,000 unique "alloy pair" entries that can be used to search for materials with tunable properties. We exemplify this approach by searching for transparent conductors and reveal candidates that might have been excluded in a traditional screening. This work lays a foundation from which materials databases can go beyond stoichiometric compounds, and approach a more realistic description of compositionally-tunable materials.

\end{abstract}

\maketitle

\section{Introduction}

\begin{figure*}
    \centering
    \includegraphics[width=\textwidth]{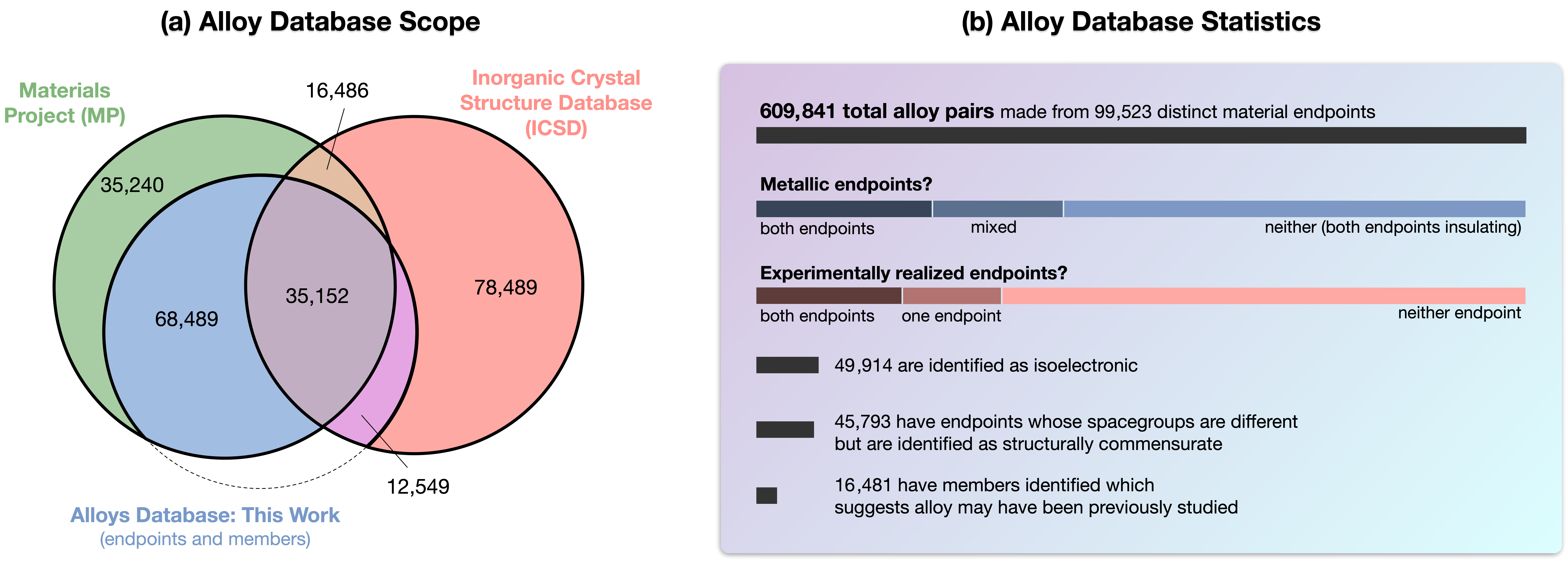}
    \caption{\textbf{Alloys database overview.} (a) A Venn diagram showing materials that are in the Materials Project (MP), associated with an alloy in the alloy database presented in this work (whether an alloy endpoint or an alloy member), and in the Inorganic Crystal Structure Database (ICSD). Note that each entry in MP represents a distinct polymorph, whereas duplicates are present in the ICSD, and so the ICSD is likely overcounted. The Venn diagram gives an overview of how these three databases relate to each other. (b) A summary of statistics within the alloy database presented in this work.}
    \label{fig:stats}
\end{figure*}

The power of functional semiconductor materials lies in the tunability of their properties. Since the dawn of the Semiconductor Age, traditional semiconductors---elemental (e.g., Si), IV-IVs (SiC), III-Vs (GaN, GaAs, InGaN), II-IVs (CdTe), etc.---have been manipulated in the laboratory through doping, alloying, processing, and other techniques to yield desired properties. Tunable semiconductor alloy materials enable a variety of energy and optoelectronic applications that govern our modern world, from light-emitting diode (LED) materials, e.g., \ce{In_{1-$x$}Ga_{$x$}N} (InGaN),\cite{mukai1999characteristics} to infrared detectors, e.g., \ce{Pb_{1-$x$}Sn_{$x$}Te} and \ce{Hg_{$x$}Cd_{1-$x$}Te}\cite{kinch2000fundamental} to piezoelectrics, e.g., \ce{PbZr_{$x$}Ti_{1-$x$}O3},\cite{anton2007review} and are critical for the transformation to renewable energy in solar cell materials, e.g., \ce{CuIn_{$x$}Ga_{1-$x$}(S_{$y$}Se_{1-$y$})2} (CIGS)\cite{dhere2006present} and \ce{CdSe_{$x$}Te_{1-$x$}} (CdTe). The properties of each of these materials reach far beyond those of their endpoint compositions, e.g., the band gap of InGaN is tunable across a wide range, from $\sim$0.7 eV (InN) to 3.4 eV (GaN). Naturally occurring semiconductor minerals are also stable in alloy forms, e.g., olivine \ce{(Mg_{$x$}Fe_{1-$x$})_2SiO4}, plagioclase \ce{Na_{$x$}Ca_{1-$x$}(Al_{$y$}Si_{1-$y$})_4O8}, and cobaltite \ce{Co_{$x$}Fe_{1-$x$}AsS}, indicating a strong tendency toward off-stoichiometric stability.\cite{rubin1997mineralogy}

Meanwhile, in the past decade computational materials discovery has been advancing novel materials design in a wide range of applications, from thermoelectrics,\cite{zhu2015computational} to Li-ion battery cathodes,\cite{chen2013sidorenkite} to transparent conductors.\cite{hautier2013identification} In most of these cases, materials discovery has been targeted towards stoichiometric ``bulk'' compounds (also called ``parent'' compounds or ``endpoint'' compounds in the context of alloys). A candidate compound emerges successfully from a screening if it satisfies a set of property values within a specific cutoff. This methodology has served as a useful starting point, but a grand challenge in the field is determining how to expand this success beyond compounds into off-stoichiometric space to search for \textit{ranges of tunability} within materials in a high-throughput context. Indeed, a material may be excluded by its endpoint properties without taking into account how its properties can be tuned by doping or alloying. For example, the n-type transparent conductor Sn-doped \ce{In2O3} is an excellent example of a material where the computed properties of the endpoint compound (\ce{In2O3}) are not representative of the high experimental performance achieved by introducing tunability.\cite{woods-robinson2018assessing}. It is recognized that considering all possible off-stoichiometry (defects, dopants, impurity phases and alloying) --- intentional as well as unintentional --- in the design of novel materials incurs a vast increase in complexity of search space compared with on-stoichiometric compound space. Therefore, part of the challenge is a data problem: how do we manage the additional complexity induced by including off-stoichiometry?

There have been many extensive and notable previous efforts to designing alloys using high-throughput computation. These include but are not limited to the design of high-entropy alloys\cite{high-entropy-allows-review, high-entropy-alloys-2}, high-entropy oxides\cite{high-entropy-oxides}, Heusler compounds\cite{heusler-review} and magnetic Heuslers \cite{heusler-magnets}, as well as alloy design for specific applications including magnetocalorics\cite{magnetocaloric-alloys} and thermoelectrics\cite{thermoelectrics}. Such design studies often bootstrap alloy searches from existing computational databases, such as the Materials Project (MP), Automated Flow for Materials Discovery (AFLOW)\cite{aflow} and the Open Quantum Materials Database (OQMD)\cite{oqmd-precipitates}. Previous efforts have also used novel approaches\cite{pareto-optimal-alloys, alloy-supercell-alternative} including machine learning\cite{ml-alloys} and density functional theory (DFT)-supported calculation of phase diagrams (CALPHAD) methodologies\cite{high-throughput-calphad}. The importance of considering alloys in high-throughput computation is therefore well known.\cite{aflow-review-alloy-mention} However, what many of these prior examples have in common is that they are often focused on the generation of new alloy materials within a limited regime of phase space; this is often from the enumeration of possibilities from a single crystal structure prototype, or is limited to binary alloys or a restricted chemical space. In contrast, our current work differs in that it offers a general approach for classifying and searching pre-existing high-throughput computational databases. These databases might already contain hidden within them sufficient information to assess the possibility of various alloys existing, but require appropriate analysis to unlock. The new analysis capabilities proposed in this work to classify and search existing databases enables more effective materials discovery screenings.

To clarify the scope of this work, we will recap what we mean by ``alloy'' in this context. The Hume-Rothery rules,\cite{mizutani2012hume} traditionally applied to metals, provide a guideline for considering whether two materials (A and B) may form a substitutional solid solution with each other (\ce{A_{$x$}B_{1-$x$}}), whereby one atom is replaced by another but the host lattice remains largely unchanged, except for small local distortions. These rules require that (1) that the crystal structures of solute and solvent must be similar (that is, commensurate with each other \cite{holder2017novel}); (2) that the atomic radius of solute and solvent atoms must differ by no more than 15\%; (3) that solvent and solute have the same valency for complete solubility; and (4) that the solute and solvent should have similar electronegativity. These rules are good guidelines, although the cutoffs (``15\%'', ``similar'') are open to debate. The methodology presented in this work therefore is focused primarily on rule (1) to generate the database using existing algorithms for assessing crystal structure similarity, with sufficient metadata then retained to assess rule (3) by querying the database. Rules (2) and (4) are easily applied by the person retrieving alloys from the database subject to their own materials design requirements; for example, by accessing the database of ionic radii within \texttt{pymatgen} to further filter down the list of possible alloys to consider. We emphasize that the alloy database obtained in this work is only a database of \emph{possible} alloys with respect to these rules, and does not guarantee that these alloys do indeed exist. Rather, it is intended as a pre-selection step to guide further inquiry.

Using this database, we create methodology to aid in the analysis of alloying opportunities, enabling computational screening for tunable properties in inorganic alloys when starting from a database of crystallographic structures and associated properties. First, we map tunable material space and search for substitutional alloy compositions and properties within a given set of possible endpoints. Second, we apply this framework to the entire MPdatabase\cite{jain2013commentary} for commensurate\cite{holder2017novel} (structure-matching within a certain tolerance; see Methodology) structures to enable analysis resulting in over 600,000 potential alloys (``alloy pairs''), encompassing 270,545 chemical systems and 215 space groups. Third, we provide a series of new techniques to conceptualize and explore this large alloy space, including defining an ``alloy system'' comprised of alloy pairs, thermodynamic stability estimates of alloys by alloy content using a "half-space hull" approach, and an example of using this data as a pre-selection step in a high-throughput screening. And lastly, we outline the limitations of this framework, and suggest next steps for tunable material screenings.

We focus on semiconductors in this paper, but the general methodology could be applied to any alloy systems where there is a reasonable expectation of structural stability and approximately linearly-dependent properties with composition. The alloy framework developed in this work is available in the open-source \texttt{pymatgen-analysis-alloys} repository and the analyses and associated enabling functionalities have been incorporated into the Materials Project website under a Creative Commons license, with an application programming interface (API) to enable other researchers to explore the data and download the results. These are online at \url{https://github.com/materialsproject/pymatgen-analysis-alloys} and \url{https://materialsproject.org/api} respectively. 
\section{Results and demonstrations}

\subsection{Creating an alloy database of "alloy pairs"}

In brief, our methods combine sets of structurally commensurate endpoint compounds into an "alloy pair" database record, which represent two compositions with the possibility of forming a solid solution with one another (see Methodology and SI for details). For example, endpoint compounds wurtzite \ce{GaN} and wurtzite \ce{InN} form an alloy pair \ce{Al_{$x$}Ga_{1-$x$}N}. 

Applying the methodology described here to the MP database produces an ``alloy database'' of 609,841 alloy endpoint pairs and 11,876 alloy systems. Of these candidate alloys, 16,481 pairs (2.7\%) and 968 systems (8.1\%) are found to contain members of intermediate, non-stoichiometric "member" compositions, suggesting that these may have been previously explored either experimentally or computationally. \textbf{\autoref{fig:stats}} depicts a summary of the dataset, as a subset of both MP\cite{jain2013commentary} and the Inorganic Crystal Structure Database (ICSD),\cite{bergerhoff1987crystallographic} and is broken down by categories including whether the alloy is metal-metal, metal-semiconductor, or semiconductor-semiconductor, and whether the alloy endpoints have been previously synthesized experimentally. While the candidate alloy pairs are generated from the MP database, alloy members are assigned by searching both the MP and ICSD databases. It is observed that of members from the ICSD, 67\% are of disordered compounds, compared to the ICSD as a whole in which 44\% are disordered. Note that exact numbers will vary according to the version of the respective database accessed, and reported statistics here reflect the most recent version of the ICSD accessible by MP at the time of publication.

In \autoref{fig:stats}, we also highlight that we have determined 45,793 alloy pairs whose endpoint compounds are not detected to have the same space group. This can either be because the detected space group, being subject to numerical tolerances, is incorrect, or it can be a sign of a phase transition. An instance of the latter case might be one endpoint of an alloy pair having a small polar distortion, while the other endpoint might be a non-polar material; here, the space groups of the endpoints do not match, but the crystal structures might still be sufficiently ``commensurate'' and able to alloy. This demonstrates the importance of carefully selecting the method for which two materials are considered to be structurally commensurate, and so might form a substitutional alloy. In the context of a materials screening, including alloys drastically expands the accessible and searchable parameter space (see Figure S3 in the SI). When properties of an alloy pair are considered, we take properties of the end-points when known and assume Vegard's law with no bowing for lattice constant, band gap ($E_\mathrm{G}$) and inverse effective mass ($1/m^*$).\cite{singh2007electronic} We note that excluding bowing is a crude approximation for band gap, but bowing is not as significant for inverse effective mass (see SI).\cite{piprek2013semiconductor}

\begin{figure*}
    \centering
    \includegraphics[width=\textwidth]{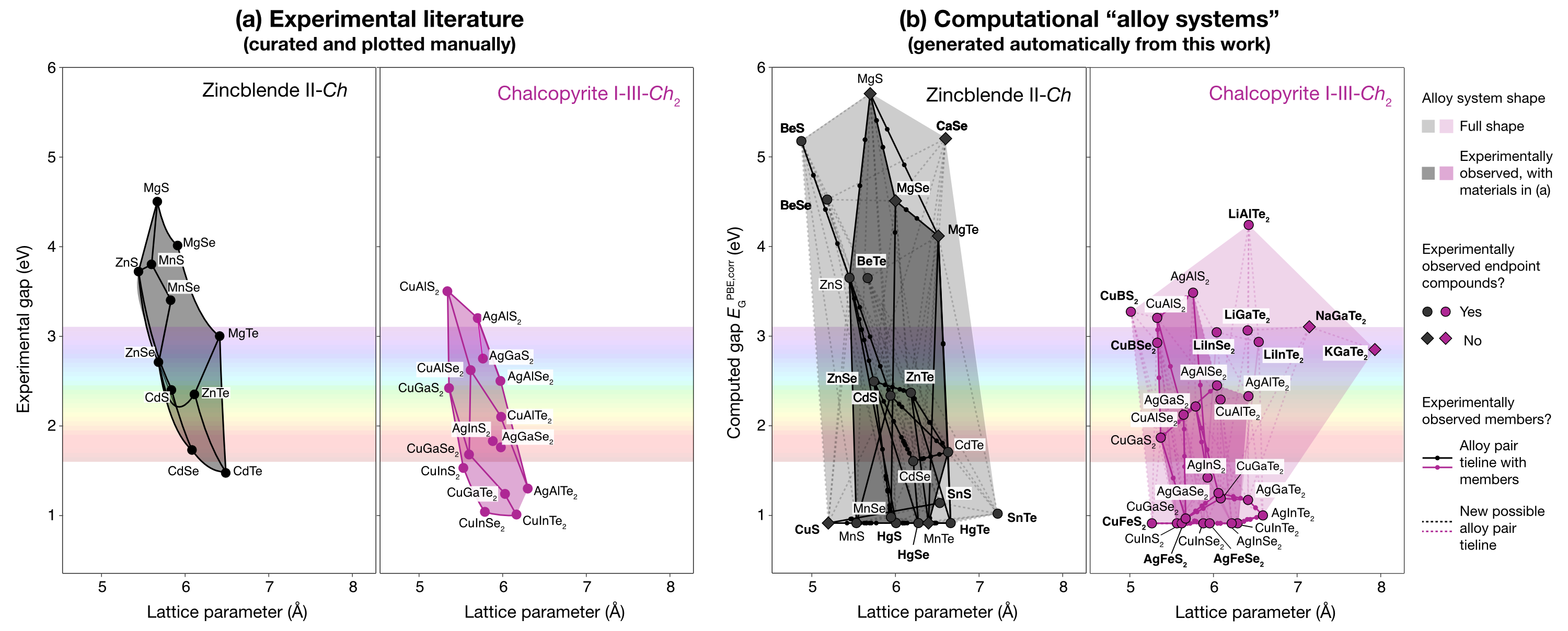}
    \caption{\textbf{Lattice parameter vs. band gap demo.} A comparison of a lattice parameter versus band gap plot for wide-band-gap zincblende and chalcopyrite chalcogenide materials (a) a manually-constructed plot from the experimental literature,\cite{woods-robinson2020wide} including experimental bowing, and (b) generated computationally from an "alloy system" in our database. Note that in (b) each alloy system is filtered to include only chalcogenide (S, Se, Te) compounds with commensurate oxidation states, that band gaps $E_\mathrm{G}^\mathrm{PBE,corr}$ are computed PBE gaps with an approximate empirical correction factor,\cite{morales2017empirical} and that the lattice parameter is computed from the conventional unit cell. Boldface in (b) indicates phases not present in (a).}
    \label{fig:alloy-system-shape}
\end{figure*}

\subsection{Exploration of alloy systems} 

By combining alloy pairs that are all commensurate with one another, ``alloy systems'' can be generated (see Methodology) in which each alloy system spans a region of accessible phase space. Applying this methodology to MP creates a total of 11,876 possible alloy systems. One application of the alloy system framework is the construction of semiconductor bowing plots, which are useful for visualizing lattice matching and band gap tuning in semiconductor alloys, and are typically constructed manually via a literature review. A typical example might be a plot showing wurtzite III-V alloys system (GaN, InN, etc.), but can be generalized for any alloy system. In \textbf{\autoref{fig:alloy-system-shape}}(a), we take an example of two systems that have been studied experimentally but not as extensively as the III-V system: zincblende II-\textit{Ch} and chalcopyrite I-III-\textit{Ch}\textsubscript{2} chalcogenide materials \cite{woods-robinson2020wide} Compounds are grouped by commensurate structure, each marker corresponding to an experimentally observed endpoint compound, and each line segment corresponds to an experimentally observed alloy (e.g., \ce{Zn_{$x$}Mg_{1-$x$}S}). Most of the compounds plotted are in their most stable polymorph; however, we note some exceptions (e.g., MnN, MnSe, and CdSe have a more stable polymorph than zincblende, but zincblende is plotted here for clarity and completeness).

Using the alloys systems framework, we generate corresponding alloy systems for zincblende and chalcopyrite chalcogenide semiconductors. These systems are plotted in the two panels of \autoref{fig:alloy-system-shape}(b) as a function of lattice parameter $a$ and band gap $E_\mathrm{G}^\mathrm{PBE,corr}$, where each commensurate system is merged into a shape to represent their range. Alloy systems are generated as a function of a single compound --- in this case zincblende ZnS and chalcopyrite \ce{CuAlS2} --- and then outputs are filtered to include only chalcogenide compounds with commensurate oxidation states. Since semi-local DFT underestimates the band gap, we plot Purdue-Berke-Ernzerhof (PBE) gaps with an applied approximate empirical correction factor from the literature, denoted as $E_\mathrm{G}^\mathrm{PBE,corr}$.\cite{morales2017empirical} Discrepancies between the experimental plots and plots derived from the database derive mainly from errors as a result of using the PBE functional. Errors in predicted gap are also exacerbated across the database in magnetic systems where the magnetic order has not been predicted; for example, in Fig 2 it is shown that zincblende MnS (Materials Project: mp-1783) is predicted with PBE to have a band gap of 0 eV, but experimentally has been shown to have a gap of approximately 3.8 eV;\cite{goede1988energy} in this case, it is because the database entry was calculated in a ferromagnetic configuration rather than the correct antiferromagnetic configuration. For better accuracy, we recommend performing additional hybrid functional calculations to complement the initial screening and provide a better estimate the gap in the alloy database or, in the future, using more accurate calculations to construct the database. We emphasize that the purpose of this work is not to demonstrate accurate band gap prediction, since more accurate methods are already well known, but to demonstrate the machinery of constructing alloy pairs and connecting these into alloy systems for the purposes of a materials discovery screening.

We observe in \autoref{fig:alloy-system-shape} that the shapes and features of computationally generated alloys systems in (b) qualitatively match the experimental diagrams in (a), subject to uncertainties in band gaps as explained above. Additionally, more information is captured in (b); in particular, the members (MP and ICSD) of many of the alloy pairs are denoted to indicate which alloys have seen previous study. Including additional hypothetical alloy pairs here increases the range of search space, by nearly 50\% percent for II-\textit{Ch} and by over 50\% for I-III-\textit{Ch}\ce{_2}, and new alloy pairs are marked with dotted lines such as \ce{Ca_$x$Cd_{1-$x$}Se} and  \ce{AgAl(Se_$x$Te_{1-$x$})2}. The computed alloy system plots can also inspire new materials design searches over a variety of multinary alloys. For example, in a search for an amber LED material ($\sim$580--590 nm; i.e., 2.10--2.14 eV) with a lattice parameter matched to zincblende GaAs (5.6531 Å\cite{haynes2016crc}), one may examine the region around \ce{Zn_{$x$}Mn_{1-$x$}S} or \ce{CuAl_{$x$}Ga_{1-$x$}Se2} alloy pairs.

In principle, for a given alloy system, an alloy's lattice parameter (or volume cube root, if comparing non-cubic systems) and band gap can be tuned within the bounds of the shape bounded by the alloy end-members in the plot by varying alloy composition. Here we show a plot for a simple comparison of $a$ vs. $E_\mathrm{G}$ for conventional semiconductors, but alloy system plots can be created for any set of properties and can in principle be expanded into higher dimensions. Some degree of bowing is likely in these systems, as shown in \autoref{fig:alloy-system-shape}(a). Additionally, discontinuities in Vegard's law can arise when gaps transition from direct to indirect nature across alloy space. However, this analysis is helpful as an initial guide and provides a systematic method for generation of such figures which are already commonplace in the literature.

\begin{figure*}
    \centering
    \includegraphics[width=140 mm]{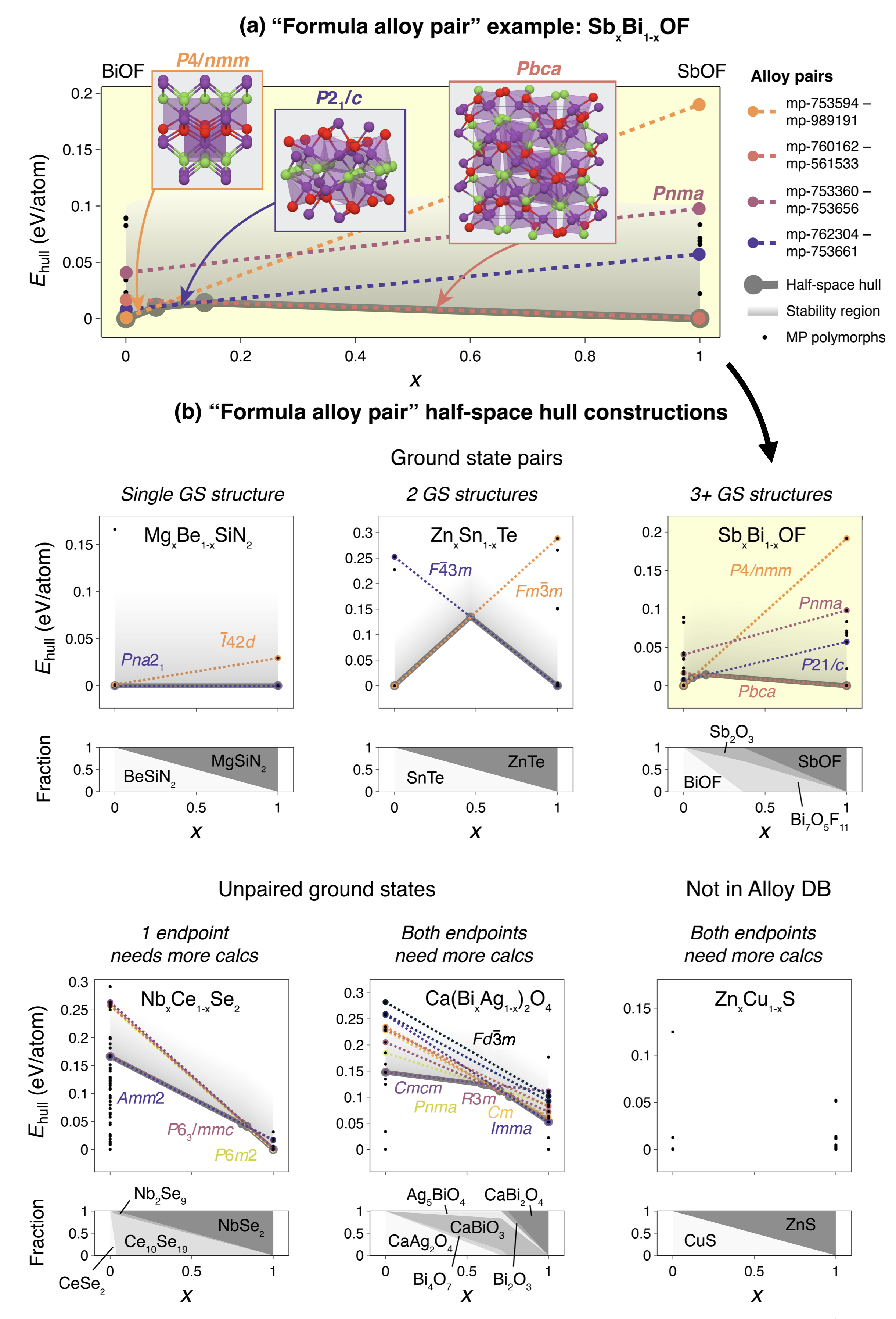}
    \caption{\textbf{Half-space hull stability analysis.} (a) A representative formula alloy pair of \ce{Sb_{$x$}Bi_{1-$x$}OF}, with three unique phases lying on the half-space hull. Dotted lines depict where a competing polymorph becomes stable. Decomposition plot on the right shows thermodynamic decomposition products from a ternary phase diagram, as a function of $x$. (b) A set of half-space hull intersections, a simple interpolation based on endpoint formation energies to find cross-overs, for six representative formula alloy pairs. "Paired" means that commensurate structures exist, while "unpaired" means that no commensurate structure is known in the database. It is not expected these cross-over points will be exact but might provide an estimate. For each alloy system, this then gives a range of allowed compositions and phases. Below each half-space hull construction, a decomposition diagram is plotted as a function of $x$.}
    \label{fig:half-space-hull}
\end{figure*}

\subsection{Estimating alloy stability}

So far we have introduced a set of alloys that are purely hypothetical, yet have not discussed thermodynamic stability and synthesizability. Computationally determining synthesizability of a given alloy is non-trivial, and requires constructing a temperature-dependent phase diagram, computing the effects of various entropy terms, and potentially considering the effects of nucleation and kinetics. A simple (yet imperfect, due to the typical $T$ = 0K approximation of DFT) metric to assess stability in solid state compound materials is the ``energy above the convex hull'' ($E_\mathrm{hull}$), where an $E_\mathrm{hull}$ of 0 eV/atom defines a thermodynamic ground state, and $E_\mathrm{hull}$ is reported for compounds in the MP database as derived from phase diagrams constructed from DFT calculations. For alloys, even determining this simple $E_\mathrm{hull}$ metric is nontrivial; it requires computing a variety of orderings, as well as fully exploring possible competing polymorph phases and all of their possible orderings (see Discussion).

To approximate whether a given alloy may be stable or synthesizable, and as a first step before performing additional in-depth calculations, we have calculated a hull across alloy content using the half-space intersection of the lines representing the linear interpolation of formation enthalpies between the two alloy endpoints --- hereafter called a ``half-space hull'' --- to identify ranges of alloy content $x$ at which different polymorphs might be stable. This is defined by a ``formula alloy pair'' and ``alloy segments'' (see Methodology). To use this database effectively, the user should make sure to consider entropic terms. Some of these, such as configurational entropy, are trivial to calculate under certain assumptions, while others such as vibrational or electronic entropy, might require further calculations or more sophisticated models.

\textbf{\autoref{fig:half-space-hull}} depicts a set of formula alloy pair diagrams derived from the alloy database, made up of alloy pairs that all have the same composition  (\ce{A_$x$B_{1-$x$}}). For example, a \ce{Sb_$x$Bi_{1-$x$}OF} formula alloy pair is magnified in \autoref{fig:half-space-hull}(a) and shown in the third panel of (b), with \ce{BiOF} endpoint compounds on the left side ($x$=0) and \ce{SbOF} endpoint compounds on the right side ($x$=1), and with the y-axis representing $E_\mathrm{hull}$. Only compounds that are present in the MP database are included here. For each case where a \ce{BiOF} compound is structurally commensurate with a \ce{SbOF} compound, an \texttt{AlloyPair} is formed and a colored dotted line is drawn in \autoref{fig:half-space-hull}(a). For example, $P4/nmm$ \ce{BiOF} (Materials Project: mp-753594, on the hull) is connected with $P4/nmm$ \ce{SbOF} (Materials Project: mp-989191, with $E_\mathrm{hull}$=0.192 eV) by a blue dotted line, while $Pcba$ \ce{BiOF} (Materials Project: mp-760162, with $E_\mathrm{hull}$=0.017 eV) is connected with $P4/nmm$ \ce{SbOF} (Materials Project: mp-561533, on the hull) by a green dotted line. In this formula alloy pair, $P2_1/c$ (purple) and $Pnma$ (red) pairs are also drawn.

The half-space hull is drawn as a continuous gray line in \autoref{fig:half-space-hull}(a). The changes of slope along this line represent ``alloy segments'' (seeMethodology), which represent phase changes as $x$ is increased. Thus, in this example the half-space hull defines segments of \ce{Sb_$x$Bi_{1-$x$}OF} where $P4/nmm$ is the lowest energy phase (0 $\lesssim x \lesssim$ 0.05), where $P2_1/c$ is the lowest energy phase (0.05 $\leq x \lesssim$ 0.15), and where $Pbca$ is the lowest energy phase (0.15 $\lesssim x \leq$ 1). Since a phase does not have to lie on the hull to be synthesizable, we draw a region above the half-space hull (the stability region in a shaded gray gradient) at which the ``energy above the half-space hull'' is less than 0.1 eV/atom. It is typical in materials screenings to define an arbitrary cutoff such as this, below which materials are more likely to be synthesizable. While this choice of cutoff is reasonable for many semiconductors, and especially oxides,\cite{aykol2018thermodynamic} we note that it would likely be far smaller for metallic alloys\cite{curtarolo2005accuracy} and may be larger for nitrides and other non-oxide semiconductors.\cite{sun2016thermodynamic} The choice of cutoff is a free parameter for the user of this database, and we encourage users to carefully consider which cutoff is most appropriate for their application.

According to the cutoff selected of 0.1 eV/atom used in \autoref{fig:half-space-hull}, it may be possible to synthesize alloys that lie within the gray region, rather than only the alloys that lie directly upon the half-space hull. For example, it may be possible to synthesize $Pbca$ \ce{Sb_$x$Bi_{1-$x$}OF} at small values of $x$, where it is not the lowest energy polymorph, because the linearly interpolated energy is still below this cutoff and close to that of competing polymorphs. However, it is far less likely that $P4/nmm$ \ce{Sb_$x$Bi_{1-$x$}OF} solid solutions could be synthesized at high values of $x$, since the interpolated energy for this alloy pair lies well outside of the 0.1 eV/atom stability region. We note that there are other endpoint compounds that do not have commensurate pairs (black circular markers), and for this method to be technically complete the formation energies of the commensurate structure pairs for these polymorphs would have to be computed. 

\begin{figure*}
    \centering
    \includegraphics[width=\textwidth]{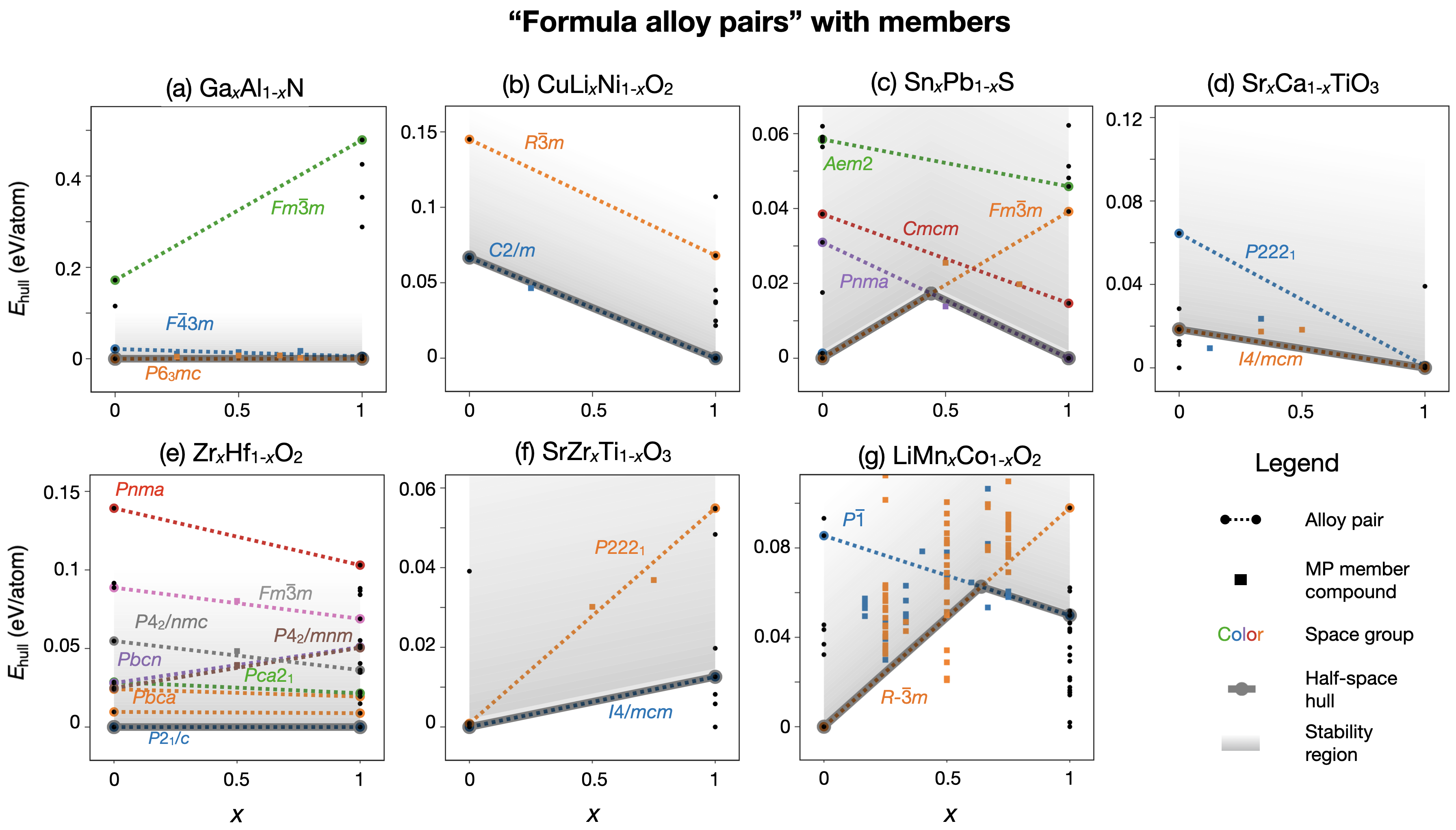}
    \caption{\textbf{Formula alloy pairs with members.} Examples of seven representative formula alloy pairs with members included. The $E_\mathrm{hull}$ of each member is sourced from the Materials Project database, and can be compared to the linearly-interpolated formation energy for each alloy pair. Alloy pairs (dashed lines) and members of alloys pairs (square markers) are colored by space group, and plots are as described in \autoref{fig:half-space-hull}.}
    \label{fig:fap-members}
\end{figure*}

\autoref{fig:half-space-hull}(b) depicts other possible scenarios of formula alloy pairs within the alloys database. Here, a “paired” ground state implies that, for the ground state of a given endpoint, a commensurate structure at the composition of the other endpoint exists, forming an “alloy pair.” An “unpaired” ground state implies that the ground state structure of one endpoint does not have a commensurate structure at the composition of the other endpoint, and this is an indication to us that more calculations should be performed in order to complete the formula alloy pair. Cases where both endpoints have paired ground states (three examples on the left) are most likely to provide useful information using the half-space hull method. For example, in \ce{Mg_$x$Be_{1-$x$}SiN2}, both ground states are $Pna2_1$ and thus it is likely that a solid solution can be synthesized across all values of $x$ with this structure retained. In \ce{Zn_$x$Sn_{1-$x$}Te}, both endpoints have commensurate ground states and no other known polymorphs with $E_\mathrm{hull}<$0.1 eV/atom. Thus a phase change from $Fm\bar{3}m$ to $F\bar{4}3m$ is expected at approximately $x$=0.5 using the half-space hull formalism. However, there are systems where one or both of the ground states do not have a commensurate pair (three examples on the right), such as \ce{Nb_$x$Ce_{1-$x$}Se2} and \ce{Ca(Bi_$x$Ag_{1-$x$})2O4}, and thus more calculations are needed in order to construct a reliable half-space hull. We note that this would add more possibly unstable or unsynthesizable endpoints.

Below each formula alloy pair in (a) is a fractional decomposition diagram. This consists of the various thermodynamic decomposition products and their fractional ratio, as a function of $x$, and is computed using existing functionality in \textit{pymatgen}. The decomposition products give an indicator of possible competing phases that may impede the formation of a solid solution, for example if a well-known material exists as a possible decomposition product, it is less likely the alloy might be synthesizable. The decompositions products are derived from the full phase diagram of the appropriate chemical system. In the \ce{Mg_{$x$}Be_{1-$x$}SiN_2} and \ce{Zn_{$x$}Sn_{1-$x$}Te}, the decomposition products consist solely of the endpoint compounds, and their relative fractional ratio increases monotonically with $x$. However the fractional decomposition of \ce{Sb_$x$Bi_{1-$x$}OF}, enlarged and plotted on the right-hand panel of (b), is more complicated and consists of four decomposition products: endpoints \ce{BiOF} and \ce{SbOF}, as well as \ce{Sb2O3} and \ce{Bi7O5F11}. Thus, although the half-space hull interpolated energies lie below 0.1 eV/atom, these \ce{Sb_$x$Bi_{1-$x$}OF} alloys may be challenging to synthesize due to competing thermodynamic reaction products.

As a check to whether the half-space hull is appropriate as a screening tool --- or in other words, whether the linearly interpolated half-space hull estimate is consistent with the DFT computed convex hull of known alloy members --- we can include members on these plots for systems in which members are present in databases and their $E_\mathrm{hull}$ values are known. For example, in \textbf{\autoref{fig:fap-members}}(a) we showcase members in the formula alloy pair construction for \ce{Ga_$x$Al_{1-$x$}N}. It is shown that the calculated formation enthalpy of the wurtzite (space group $P6_3mc$) alloy members lie below the zincblende (space group $F\bar{4}3m$), which is consistent with with the half-space hull; here, these data points refer to the formation enthalpies as calculated with DFT using small ordered approximations from entries already existing in the MP database. In \autoref{fig:fap-members}(b--g), we plot six other examples of formula alloy pair half-space hull constructions for which there are members included in the alloy pairs. For \ce{Zr_$x$Hf_{1-$x$}O2} (e), the calculated formation energies for space groups $P4_2/mnm$, $P4_2/nmc$, and $Fm\bar{3}m$ at $x$=0.5 lie nearly exactly on the linearly-interpolated energies. Other systems (e.g., \ce{Sr_$x$Ca_{1-$x$}TiO3} and \ce{Sn_$x$Pb_{1-$x$}S}) have alloys ranked in the same order as the half-space hull prediction, albeit not precisely on the predicted lines. We note that some ``alloys'' are extensively sampled in the MP database such as \ce{LiMn_$x$Co_{1-$x$}O2}, likely due to its interest as a battery material leading to a large amount of calculations performed on this compound with varying degrees of lithiation. All of these plots are generated using the tools provided by \texttt{pymatgen-analysis-alloys}, and can be similarly constructed for any system of interest. We also perform a simple statistical analysis on the full set of formula alloy pairs with members, and find this framework yields correct polymorph orderings for a majority of the set (to within 25 meV/atom error for 64\% of set, and to within 100 meV/atom error for 91\% of the set; see Figure S1 in the SI).

Therefore, when analyzing the set of alloy pairs or alloy systems within our database, it is important to assess which half-space hull scenario a given formula alloy pair lies within and whether there exists a region of phase space where stabilization is likely. Additionally, assessing possible decomposition products informs whether to expect multiple decomposition products, which could impede formation of the alloy. Overall, the half-space hull framework of drawing lines to estimate segments of phase stability is not rigorous, since formation enthalpy does not follow Vegard's law and configurational entropy is not taken into account. Rather, this method is intended to provide an estimate of what alloys might be present and where in alloy space they might be, as a tool to justify or prioritize additional calculations in a high-throughput context, and it is therefore an entry-point for determining which alloys may be experimentally realizable.

\begin{figure*}
    \centering
    \includegraphics[width=\textwidth]{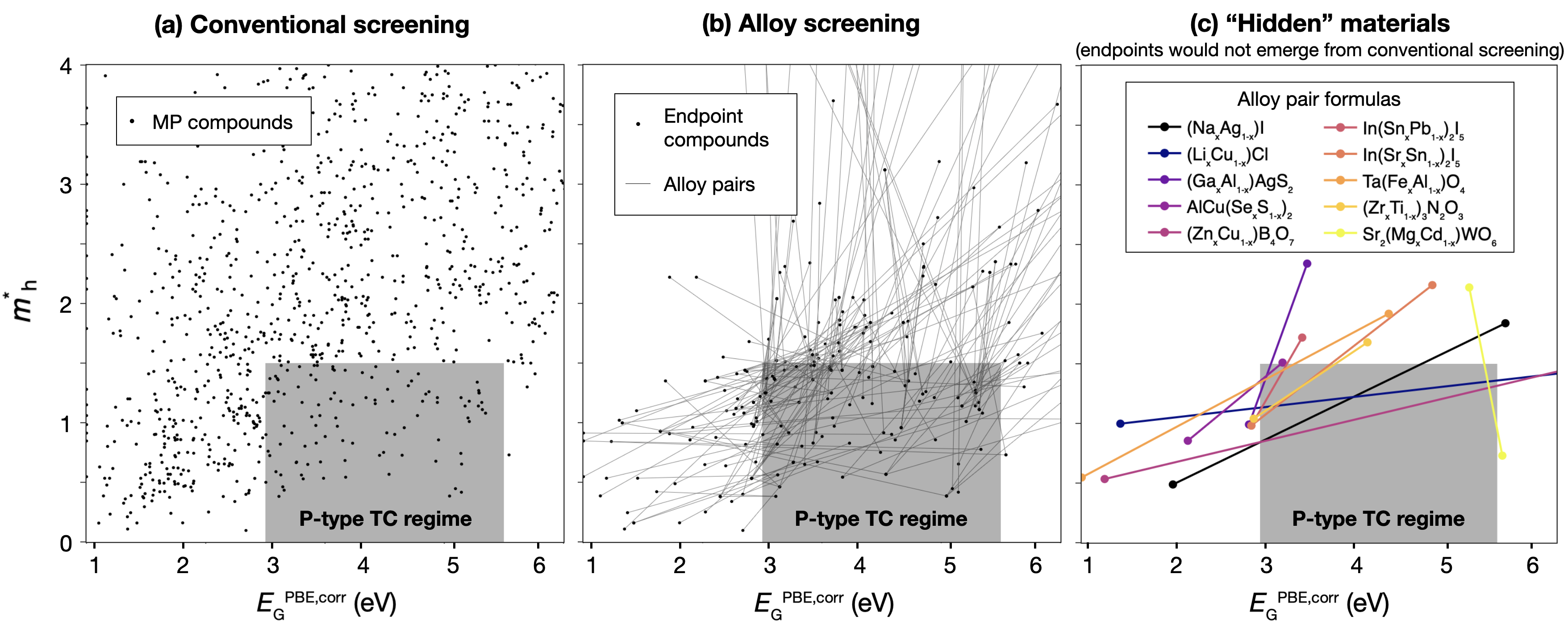}
    \caption{\textbf{Example screening for p-type transparent conductors (TCs).} An example of a computational screening of an alloy search space, with the approximate computed p-type TC regime designated with a gray box. (a) All bulk compounds that intersect the approximate p-type TC regime. (b) All alloy pairs that intersect the approximate computed p-type TC regime. (c) ``Hidden'' alloy pairs that intersect the p-type TC regime where both endpoints lie outside of the regime. Pairs are denoted by the range of their fractional alloy compositions which lie within the regime, with details denoted in \autoref{tab:p-TC-alloys}.}
    \label{fig:p-TC-alloy-pairs}
\end{figure*}

\subsection{Example of screening alloy pairs for p-type transparent conductors}

Including alloys can expand the number of material candidates generated by high-throughput screenings, and reveal candidates that otherwise would not have emerged. Here, to demonstrate this quantitatively, we screen our candidate alloy pair dataset for possible p-type transparent conductor (TC) candidates. Discovery of a high-performance p-type TC could enable breakthroughs in solar cells and transparent electronics, among other applications, but to date there are no p-type TCs that perform as well as n-type TCs.\cite{banerjee2005recent} A high-performing p-type TC is likely to require a low hole effective mass ($m_\mathrm{h}^*$) to enable high hole mobility and a wide band gap ($E_\mathrm{G}$) to enable optical transparency, among other properties.\cite{woods-robinson2018assessing} So far, several data-driven explorations have been performed to search for p-type TC candidates,\cite{hautier2013identification, sarmadian2016easily, varley2017descriptorbased} but to our knowledge no screenings have been performed looking specifically for alloys or specifically for tunable materials rather than compounds.

For this analysis, we use a set of stable or metastable ($E_\mathrm{hull} < $ 0.1 eV) representative compounds (i.e. alloy endpoints) where $m_\mathrm{h}^*$ has been calculated (the same set shown in Figure S3).\cite{ricci2017abinitio} First, \textbf{\autoref{fig:p-TC-alloy-pairs}}(a) shows a set of bulk compounds from the MP database, with empirically-corrected band gap\cite{morales2017empirical} $E_\mathrm{G}^\mathrm{PBE,corr}$ on the x-axis and $m_\mathrm{h}^*$ on the y-axis. The gray ``p-type TC regime'' depicts a range of parameter space where 2.9 eV $< E_\mathrm{G}^\mathrm{PBE,corr} <$ 5.6 eV (and thus likely to be transparent) and $m_\mathrm{h}^* < 1.5$, where p-type TC candidates may reside.\cite{woods-robinson2018assessing} This figure plots approximately 1,000 compounds, approximately 150 of which lie within the p-type TC region and contain compounds that have emerged from previous screenings (e.g., \ce{ZrOS}, \ce{TaCu3S4}, and \ce{Al2ZnTe4}). The choice of cutoff value tends to be motivated by expected values of physical parameters (e.g., absorption edge and hole mobility), but incur uncertainties in calculated value and inconsistencies between descriptor value and real physical value. Hence, the goal is to suggest a list of target candidates that may be suitable to prioritize for future computational study and experimental inquiry. Therefore, \textbf{\autoref{fig:p-TC-alloy-pairs}}(a) represents a conventional materials discovery screening.

In contrast, \autoref{fig:p-TC-alloy-pairs}(b) depicts a subset of alloy endpoint compounds (black circular markers) and corresponding linearly-interpolated alloy pair properties assuming Vegard's law (thin lines between points). This analysis yields 233 alloy pairs whose lines intersect the ``p-type TC regime,'' and a subset of 192 alloy pairs in which one or more endpoint lies outside the regime are plotted here for readability. Thus, this plot demonstrates a set of possible, additional alloy pairs to consider as p-type TCs that previously may have been overlooked. The alloy pairs present within the gray region indicate there may be combinations of $E_\mathrm{G}^\mathrm{PBE,corr}$ and $m_\mathrm{h}^*$ beyond those represented by the endpoint compounds in \autoref{fig:p-TC-alloy-pairs}(a).

In \autoref{fig:p-TC-alloy-pairs}(c), we take this a step further by highlighting a subset of 10 ``hidden'' alloy pairs that intersect this p-type TC regime but where both of the endpoints lie outside of the regime. This analysis illustrates compounds that themselves are not p-type TC candidates but whose alloys may warrant further exploration. \autoref{tab:p-TC-alloys} reports all the hidden pairs from this analysis, including the 10 hidden pairs from \autoref{fig:p-TC-alloy-pairs}(c). Included in this table the range of $x$ where properties lie within the p-type TC regime (``$x$ range''), the range of $E_\mathrm{G}^\mathrm{PBE,corr}$ and $m_\mathrm{h}^*$ achieved within this window, and $E_\mathrm{hull}$ of the endpoints (where $E_\mathrm{hull}^\mathrm{A}$ corresponds to the first compound of a pair and $E_\mathrm{hull}^\mathrm{B}$ to the second). It is also denoted whether a region of the $x$ range lies on the half-space hull, and the number of decomposition products (excluding the endpoint compounds from the count). Most of the alloy pairs that emerge from this screening are quaternaries (alloys of two ternary compounds; e.g., \ce{AlCuS_{$x$}Se_{1-$x$}}), with several ternaries (alloys of binary compounds; e.g., \ce{Cu_{$x$}Li_{1-$x$}Cl}) and quinternaries (alloys of quaternary compounds; e.g., \ce{Sr_{2}Mg_{$x$}Cd_{1-$x$}WO_{6}}). To our knowledge, none of these alloy pairs have been studied previously as p-type TCs, with the exception of \ce{La2SeO2} and \ce{Gd2SeO2}, which have been predicted previously using a high-throughput approach.\cite{sarmadian2016easily} We note that this is just one example of an application where including alloying could yield new material candidates.

\begin{table*}
\centering
\caption{``Hidden'' alloy pairs with properties of interest to p-type TCs.}
\label{tab:p-TC-alloys}
\setlength{\tabcolsep}{2mm}
\resizebox{\textwidth}{!}{%
\begin{tabular}{cccccccccccc}

\toprule
    \textbf{\begin{tabular}[c]{@{}c@{}} Pair IDs \\ (A--B)\end{tabular}} &
    \textbf{Alloy formula} &
    \textbf{\begin{tabular}[c]{@{}c@{}} Space \\ group\end{tabular}} &
    $\bm{x}$ \textbf{range} &            
    \textbf{\begin{tabular}[c]{@{}c@{}} $\bm{E_\mathrm{G}^\mathrm{PBE}}$ range \\ (eV)\end{tabular}} &         
    $\bm{m_\mathrm{h}}^*$ \textbf{range} &  \textbf{\begin{tabular}[c]{@{}c@{}} $\bm{E_\mathrm{hull}^\mathrm{A}}$ \\ (eV/at.)\end{tabular}} &
    \textbf{\begin{tabular}[c]{@{}c@{}} $\bm{E_\mathrm{hull}^\mathrm{B}}$ \\ (eV/at.)\end{tabular}} &

    \textbf{\begin{tabular}[c]{@{}c@{}} \textbf{On half-}  \\ \textbf{space hull?}\textsuperscript{†}\end{tabular}} &
    
    \textbf{\begin{tabular}[c]{@{}c@{}} \textbf{\# decomp.}  \\ \textbf{products}\textsuperscript{‡}\end{tabular}}
    
    \\
\midrule
\midrule

  mp-22919--mp-23268 &         
  \ce{(Na_{$x$}Ag_{1-$x$})I} &  
  $Fm\bar{3}m$ & 
  0.27--0.74 & 1.52--2.85 & 0.85--1.49 &     0.093 &     0.000 
  & yes & 0
  
  \\
  mp-571386--mp-22905 &        
  \ce{(Li_{$x$}Cu_{1-$x$})Cl} & 
  $Fm\bar{3}m$ & 
  0.20--0.52 & 1.51--3.46 & 1.13--1.36 &     0.178 &     0.020 
  & no & 1
  
  \\
  mp-684712--mp-32891 &        \ce{(Y_{$x$}Gd_{1-$x$})_{2}S_{3}} &         
  $I\bar{4}2d$ &
  0.54--0.57 &  1.50--1.53 &  1.49--1.50 &     0.022 &     0.036
  & no & 0
  
  \\
  mp-5782--mp-556916 &      
  \ce{(Ga_{$x$}Al_{1-$x$})AgS_{2}} &         
  $I\bar{4}2d$ &
  0.62--0.81 &  1.50--1.59 &  1.24--1.50 &     0.000 &     0.003 
  & yes & 0 
  
  \\
  mp-4979--mp-8016 &      
  \ce{AlCu(Se_{$x$}S_{1-$x$})_{2}} &         
  $I\bar{4}2d$ &
  0.02--0.24 &  1.50--1.68 &  1.35--1.50 &     0.000 &     0.000 & yes & 0
  
  \\
  mp-756317--mp-3536 &     \ce{Al_{2}(Mg_{$x$}Hg_{1-$x$})O_{4}} &         
  $P4/mbm$ &
  0.05--0.16 & 1.53--1.94 & 1.21--1.49 &     0.087 &     0.000 
  & yes & 1
  
  \\
  mp-756317--mp-2908 &     \ce{Al_{2}(Zn_{$x$}Hg_{1-$x$})O_{4}} &         
  $P4/mbm$ &
  0.07--0.62 & 1.52--2.91 &  1.13--1.50 &     0.087 &     0.000 
  & yes & 1
  
  \\
  mp-9081--mp-11742 &      
  \ce{CsNd(Te_{$x$}S_{1-$x$})_{2}} &         
  $R\bar{3}m$ &
  0.55--0.75 &  1.50--1.67 & 1.36--1.49 &     0.002 &     0.000 
  & yes & 0
  
  \\
  mp-555093--mp-558690 &      \ce{(Zn_{$x$}Cu_{1-$x$})B_{4}O_{7}} &          
  $Cmcm$ &
  0.26--0.63 & 1.53--3.48 & 0.85--1.31 &     0.047 &     0.058 
  & yes & 1
  
  \\
  mp-13973--mp-7233 &     \ce{(La_{$x$}Gd_{1-$x$})_{2}SeO_{2}} &         
  $P\bar{3}m1$ &
  0.14--0.15 &  1.50--1.51 &   1.50--1.50 &     0.000 &     0.000 
  & yes & 0
  
  \\
  mp-23520--mp-23417 &     \ce{In(Sn_{$x$}Pb_{1-$x$})_{2}I_{5}} &         
  $I4/mcm$ &
  0.31--0.83 &  1.50--1.73 & 1.11--1.49 &     0.056 &     0.023 
  & yes & 2
  
  \\
  mp-23417--mp-23504 &     \ce{In(Sr_{$x$}Sn_{1-$x$})_{2}I_{5}} &         
  $I4/mcm$ &
  0.05--0.44 &  1.50--2.09 &  1.04--1.50 &     0.023 &     0.046 
  & yes & 2
  
  \\
  mp-754818--mp-756933 &      \ce{(Tl_{$x$}Na_{1-$x$})TaO_{3}} &         
  $P4/mbm$ &
  0.36--0.53 &  1.50--1.91 &  1.39--1.50 &     0.087 &     0.002 
  & yes & 0
  
  \\
  mp-7482--mp-8402 &      
  \ce{Rb(Mg_{$x$}Hg_{1-$x$})F_{3}}  &         
  $Pm\bar{3}m$ &
  0.14--0.16 & 1.52--1.64 & 1.41--1.47 & 0.000 &     0.002 
  & yes & 0
  
  \\
  mp-760396--mp-761390 &      \ce{Ta(Fe_{$x$}Al_{1-$x$})O_{4}} &         
  $I4_1md$ &
  0.31--0.42 & 1.51--1.79 &  1.35--1.50 &     0.056 &     0.019 
  & no & 0
  
  \\
  mp-755054--mp-755998 &     \ce{(Zr_{$x$}Ti_{1-$x$})_{3}N_{2}O_{3}} &          
  $Cmcm$ &
  0.06--0.71 & 1.51--2.13 &  1.08--1.50 &     0.008 &     0.002 
  & no & 4
  
  \\
  mp-760655--mp-757905 & \ce{Li_{3}(Ti_{$x$}Bi_{1-$x$})(PO_{4})_{2}} &          
  $C2/m$ &
  0.44--0.60 & 1.51--2.08 & 1.16--1.48 &     0.066 &     0.072 
  & yes & 6
  
  \\
  mp-18903--mp-18848 &    \ce{Sr_{2}(Mg_{$x$}Cd_{1-$x$})WO_{6}} &         
  $Fm\bar{3}m$ & 
  0.16--0.54 &  3.39--3.50 &  0.95--1.50 &     0.082 &     0.009 
  & no & 0
  
  \\
  mp-18848--mp-19400 &    \ce{Sr_{2}(Ni_{$x$}Mg_{1-$x$})WO_{6}} &         
  $Fm\bar{3}m$ & 
  0.5--0.53 & 1.52--1.65 &  1.45--1.50 &     0.009 &     0.010 
  & no & 0
  \\
  
\bottomrule
\end{tabular}
}

\parbox[t]{\textwidth}{\centering \footnotesize \smallskip
  \textsuperscript{†}Whether a composition within $x$ range lies on the half-space hull.
  \textsuperscript{‡}Number of decomposition products from half-space hull; excludes endpoint compounds from count. 
}

\end{table*}

\section{Discussion}

We have demonstrated a framework to propose new alloys and access the potential tunability of materials for high throughput screenings. In our presented database, we designate “alloy pairs” between commensurate endpoint structures; although we present 600,000 unique pairs, this database comprises a subset of possible physical alloys. Several extensions of the presented alloy database are possible, beyond constructing structure-matched pairs. For example, in many experimentally observed alloy systems, endpoints may not structure match within the tolerances we use here but are still “commensurate” with one another, i.e. they can be connected through a displacive phase transformation (e.g., orthorhombic SnS and rocksalt CaS).\cite{holder2017novel} Such pairs are not included in this database, however, advances in methodologies for determining whether displacive phase transformations are possible between a given pair of materials could allow the database to be expanded in future.\cite{stevanovic2018predicting, therrien2020matching} In some cases incommensurate structures, where symmetries are distinct from one another but can be connected through a reconstructive transformation, can also form heterostructural alloys which are of increased interest for materials design (e.g., rocksalt MnO and wurtzite ZnO can alloy to form \ce{Mn_{$x$}Zn_{1-$x$}O}).\cite{holder2017novel} Similarly, a material might be tuned by varying vacancy concentration topotactically (e.g., \ce{NiO_$x$}). Furthermore, there are alloy pairs and alloy systems that in principle could alloy, but have no commensurate endpoint structures currently on MP (e.g., formula alloy pairs labeled ``unpaired ground states'' and ``not in DB'' in \autoref{fig:half-space-hull}, so in these systems more calculations would be required before the alloy could be defined. Nevertheless, in principle, the methodology presented here could be expanded upon to include and categorize all plausible commensurate and incommensurate alloy pairs, and each of the cases mentioned here could be incorporated into future iterations of this alloys database.

We note that the underlying, input database from which our alloy database is derived can contain biases. These biases, e.g., concerning structural as well as chemical coverage, can propagate into the alloy database, which should be acknowledged when interpreting results. For example, the MP database necessarily contains many materials that might alloy with each other due to its use of specific structure prediction\cite{hautier2010finding,hautier2011data} methodologies. As the underlying database expands, this infrastructure has been established to automatically ``build'' new versions of the alloy database as new data becomes available. Importantly, as better methods for calculating more accurate lattice parameters or band gaps become accessible for high-throughput computation, the alloy database will also incorporate this improved data. The continued building of new versions of this database is an essential aspect of this work, since static datasets have limited utility given the pace of improvement of computational materials databases.

\begin{figure*}
    \centering
    \includegraphics[width=0.61\textwidth]{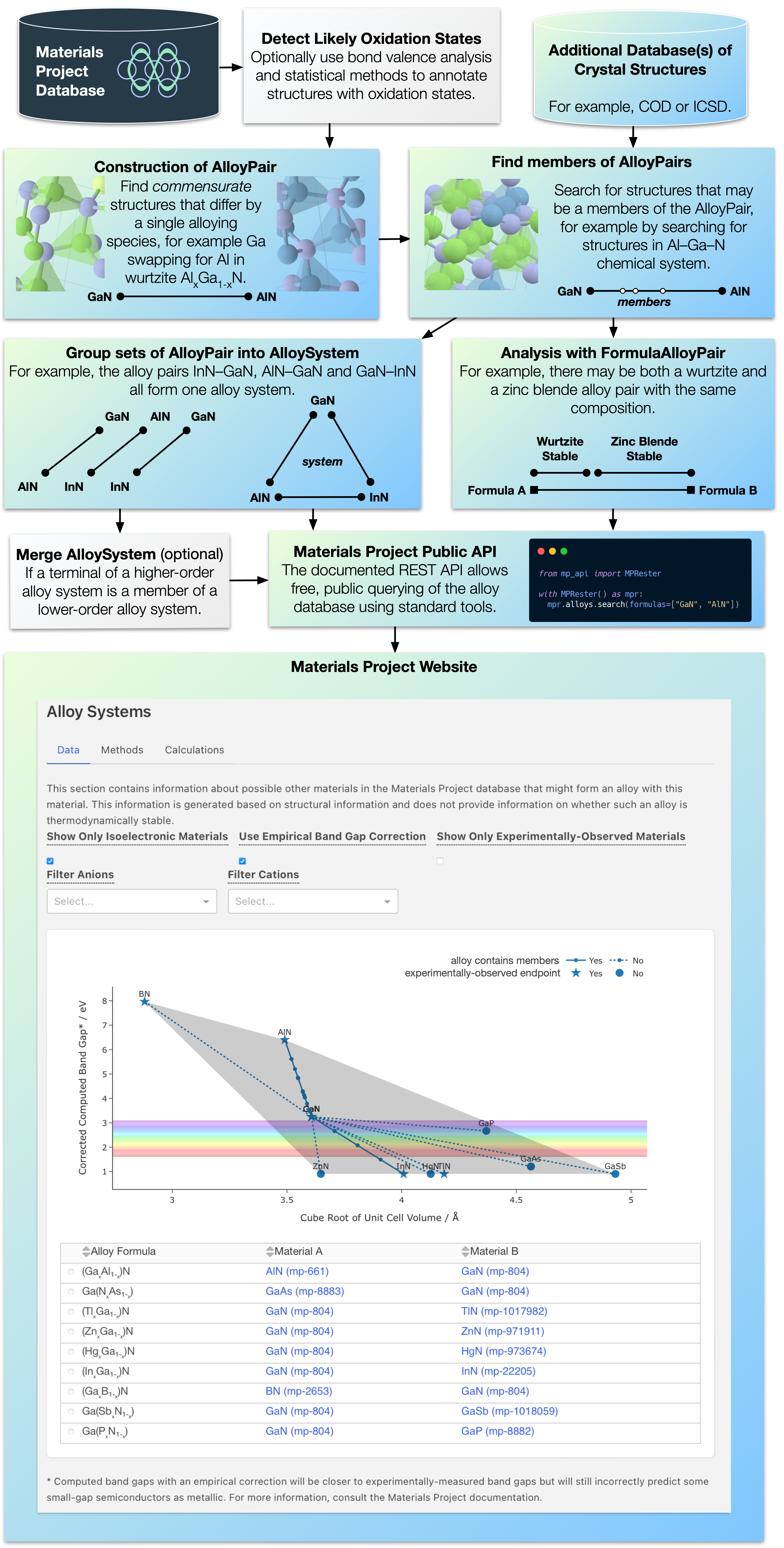}
    \caption{\textbf{Alloys data processing pipeline.} A flowchart showing the data processing pipeline outlined in the Methodology, starting from a generic crystal structure database such as the Materials Project, and ending with a publicly accessible API and website to explore the data. Wurtzite GaN is shown here as an example, as an alloy pair with InN and as an alloys system plotted on the MP website.}
    \label{fig:flowchart}
\end{figure*}

Once a set of potential alloys are suggested from this database, more reliable methods to assess alloy solubility can be used to either rule out or confirm a potential alloy; for example, automated cluster expansions\cite{laks1992efficient,barroso2022smol} or the generalized quasi-chemical approximation (GQCA) method.\cite{chen1995semiconductor} Our work is intended to serve as a starting point from which to determine systems to consider for such in-depth analyses. The half-space hull diagrams provide a guide to select alloys within a given chemical space which may be stable and synthesizable. For example, the following calculations of increasing computational cost could be explored based on outputs from the alloys database:

\begin{itemize}
    \item For compounds at endpoint A (or B) in which a commensurate compound at endpoint B (or A) is not present on MP, there is insufficient information in the database to calculate an alloy pair (for example, the black circular markers in \autoref{fig:half-space-hull} without any connecting line) such as \ce{Zn_{$x$}Cu_{1-$x$}S}). Here, the missing compound(s) can be calculated and added to the database. This is still important even if such a compound is unstable or not experimentally realizable at the endpoint, since there may be a region within alloy space where synthesizability becomes possible.
    \item For alloy pairs in which member compounds are not yet known to exist, members can be calculated (e.g., at $x$=0.5) for a few different orderings to assess realizability, or give an indication of expected bowing and other parameters.
    \item Many real alloy materials are \textit{disordered}, rather than ordered. For members within in alloy pair, special quasi-random structure (SQS) calculations can approximate structures of fully random alloy polymorphs to provide a counterpoint to the small-cell ordered structures more typical in a database such as MP.\cite{zunger1990special}
    \item To account for configurational entropy and thermodynamics of specific alloy members, the generalized quasi-chemical approximation (GQCA) can be used to estimate free energy,\cite{chen1995semiconductor} and subsequently higher order methods such as cluster expansions can be applied to further investigate specific systems for which high quality phase diagrams are required.\cite{laks1992efficient}
\end{itemize}

For immediate use, our alloy database has been incorporated into MP as an app in the new website release and API, in the hope that this will serve as a guide for researchers performing screenings of tunable materials. The alloys database will be updated alongside the MP database. A flowchart of the alloys database pipeline and incorporation onto MP is shown in \textbf{\autoref{fig:flowchart}}.

\section{Conclusion}

In this paper we have presented a new framework to analyze alloys in the context of materials databases, implemented it into the open source \texttt{pymatgen-analysis-alloys} package, and created an open-source alloys database that has been incorporated into the Materials Project website. We have presented a few case studies here of how this database can be utilized in the context of materials research and design.

Importantly, all the analysis presented here has been performed without any new calculations, which showcases some of the data analysis opportunities from mining existing databases. A decade into the Materials Genome Initiative, the materials discovery community has produced large quantities of data in multiple databases, but data \textit{production} is just the start; it is essential that data is curated, structured, and connected in a way to yield the maximum value to the community.

In particular, one of the key challenges is how to link and apply this data to successfully use computational predictions to inform experimental results, especially as experimental databases grow.\cite{zakutayev2018open,talley2021research}  In particular, experimental progress in semiconductors typically starts from a well-studied, well-characterized material and modifies its properties iteratively with the addition of dopants or alloying elements during growth. The framework of this paper addresses this aspect of materials design by creating a database of candidate, tunable materials by a data-focused approach which can use existing materials databases to suggest alloys between pairs of already-known materials. Thus, a new materials screening procedure is now possible that can emphasize experimentally-accessible materials and suggest screening outputs that would have been previously wholly overlooked.

\section{Experimental procedures}

\subsection{Resource availability}

\subsubsection{Lead contact}

Requests for information and resources should be directed to the lead contact, Dr. Rachel Woods-Robinson (rwoodsrobinson@berkeley.edu).

\subsubsection{Materials availability}

This study did not generate any chemical reagents.

\subsection{Methodology}

We have created an open-source code, \texttt{pymatgen-analysis-alloys}, that allows the construction of an alloy database when provided with an input database containing crystal structures. As a demonstration, we apply this code to the MP database. The left side of \autoref{fig:flowchart} depicts the data processing pipeline of the alloys database, as described here. This code is also used for the automatic generation of the plots shown in this manuscript, with only light additional editing performed for presentation.

The method outlined here does not require any prior knowledge of which materials might form alloys. While partial occupancies in e.g., a Crystallographic Information File (.cif) indicates the possibility of alloying, this criteria only captures known systems, and hence does not fully explore the possible alloy space.  The challenge when constructing the database is in the data processing pipeline, and addressing combinatorial problems when large databases of hundreds of thousands of entries are used.

The method is as follows: for each crystal structure in the input database, designated as a potential ``endpoint,'' we find all other compounds that share its anonymous formula (e.g., ``\ce{ABC2}''), and perform a pairwise comparison between all materials to detect commensurate structures using the \texttt{StructureMatcher} in \texttt{pymatgen}\cite{ong2013python}. A pre-filter is performed that checks for detected space group, calculated with \texttt{spglib}\cite{togo2018spglib}, using both tight and loose tolerances. This pre-filter is imposed with the logic that it is a necessary but not sufficient condition that two commensurate crystal structures will have the same space group. After a pair of crystal structures are identified as an endpoint, information is extracted such as the alloying species, including oxidation state, and whether the alloy is isoelectronic, and stored as an instance of an \texttt{AlloyPair} class. This definition of ``alloy'' does not consider alloys formed through interstitial alloying additions or other types of alloys. We apply the definition of ``commensurate'' from Holder et al.: "alloys between materials with commensurate lattices...have symmetries that are related by a displacive phase transformation" such that "structural distortions and atomic displacements lead to a continuous change of the lattice parameters and site positions without requiring atomic diffusion or rearrangement of the coordination environment."\cite{holder2017novel} Our alloy database will capture these commensurate examples if the displacements are small (such that the \texttt{StructureMatcher} algorithm considers them to be the same structure within a given tolerance) but not if displacements are large.

All \texttt{AlloyPair} entries contain structural properties, such as space group and primitive cell volume, but can be supplemented with additional properties. For this demonstration, supplemental properties are taken from the MP and include $E_\mathrm{hull}$, $E_\mathrm{G}$ from the Purdue-Berke-Ernzerhof (PBE) functional from the MP, and electron and hole effective masses from Ricci et al.\cite{ricci2017abinitio} (note that these are only computed for a subset of the MP database), but in principle this can be expanded to include any material property. Methods are provided to interpolate these properties using Vegard's law (assuming no bowing) for a given alloy content to allow for easier plotting and searching (see SI).

Once a set of \texttt{AlloyPair} entries are constructed, they are grouped by chemical system and iterated over to search for potential members, defined by the \texttt{AlloyMember} class, using a similar approach. Determining whether a provided crystal structure can be assigned as a member to a given alloy pair requires both a exact match of stoichiometry and, in addition, either an exact match of space group or a structure match. This stricter criteria reduces the total number of members that might be assigned and could be relaxed for some applications; for example, if a small off-stochiometry was allowable. Assigning membership allows a database query to reveal which alloys already have existing data available, and thus may be more experimentally accessible alloys when performing a screening.

A set of alloy pairs can be grouped together as an alloy system, defined by the class \texttt{AlloySystem}, using a network graph method whereby each edge in the graph is an alloy pair and connected subgraphs form the respective alloy systems. The code allows for alloy systems to be merged when a member of one system might be the endpoint of another system, for example an alloy system with ternary endpoints where one endpoint is itself a member of a binary alloy pair.

Another useful grouping is the set of alloy pairs that all have the same set of endpoint formulae: these can be grouped together as a formula alloy pair, defined by the class \texttt{FormulaAlloyPair}. If formation enthalpies are known for the endpoints, this class is able to define regions where a given polymorph is stable according to a simple linear interpolation, and define alloy segments (class \texttt{AlloySegment}) that encode the critical alloy contents at which a phase transition may occur between two polymorphs. Furthermore, if any alloy members are known, including their formation enthalpies, this data will  inform how accurate the simple linear interpolation may be. Examples of these can be seen in \autoref{fig:fap-members}.

For the example database generated in this work, we exclude compounds including H, He, noble gases, and heavy elements with atomic numbers greater than 83 (Bi), although all these entries are present in the underlying database, but we do not perform any further filtering based on chemistry and leave this as a capability for the user querying the database to decide exactly what chemical systems, maximum electronegativity differences, etc. are allowable for their specific design case.

The API to access and search the database is defined in the open-source \texttt{emmet} code. The user interface on the MP website is constructed using the open-source Crystal Toolkit web framework. All open-source code described in this work is open to review and suggested edits by other researchers, and any contributions are welcomed by the authors.

\section{Data and code availability}

The alloy framework developed in this work is available in the open-source \texttt{pymatgen-analysis-alloys} repository and the analyses and associated enabling functionalities have been incorporated into the MP website under a Creative Commons license, with an API to enable other researchers to explore the data and download the results (see Video S1. Exploring alloys on the Materials Project). These are online at \url{https://github.com/materialsproject/pymatgen-analysis-alloys} and \url{https://materialsproject.org/api} respectively. In addition, a static snapshot of the latest version of the data- base at the time of publication has also been made available at Figshare: \url{10. 6084/m9.figshare.22491793}. Note that the database presented in this work is a living resource and will be updated and revised over time to include additional data and fixes where applicable, so data retrieval via the API for any follow-up research purposes is strongly recommended.

\section*{Acknowledgments}

This work was supported by the U.S. Department of Energy, Office of Science, Office of Basic Energy Sciences, Materials Sciences and Engineering Division under Contract No. DE-AC02-05-CH11231 (Materials Project program KC23MP). R.W.R. acknowledges financial support from the U.C. Berkeley Chancellor's Fellowship and the National Science Foundation (NSF) Graduate Research Fellowship under Grant No. DGE1106400 and DGE175814.

\section{Author contributions}

We highlight the author contributions to this study using the CRediT taxonomy.

\textbf{R.W.R}: Conceptualization, Methodology, Investigation, Data Curation, Formal Analysis, Software, Validation, Visualization, Writing – Original Draft, Writing – Review \& Editing.

\textbf{M.K.H.}: Conceptualization, Methodology, Investigation, Data Curation, Formal Analysis, Software, Validation, Visualization, Writing – Original Draft, Writing – Review \& Editing.

\textbf{K.A.P.}: Funding Acquisition, Project Administration, Resources, Supervision, Writing – Review \& Editing.

\section{Declaration of interests}

The authors declare no competing interests.

\bibliographystyle{numbered.bst}
\bibliography{main.bib}

\end{document}